# Observation of twist-induced geometric phases and inhibition of optical tunneling via Aharonov-Bohm effects

Midya Parto, Helena Lopez-Aviles, Jose E. Antonio-Lopez, Mercedeh Khajavikhan, Rodrigo Amezcua-Correa, and Demetrios N. Christodoulides

*CREOL, The College of Optics & Photonics, University of Central Florida, Orlando, FL 32816, USA*

**Geometric phases appear ubiquitously in many and diverse areas of physical sciences, ranging from classical and molecular dynamics to quantum mechanics and solid-state physics. In the realm of optics, similar phenomena are known to emerge in the form of a Pancharatnam-Berry phase whenever the polarization state traces a closed contour on the Poincaré sphere. While this class of geometric phases has been extensively investigated in both free-space and guided wave systems, the observation of similar effects in photon-tunneling arrangements has so far remained largely unexplored. Here, for the first time, we experimentally demonstrate that the tunneling or coupling process in a twisted multi-core fiber system can display a chiral geometric phase accumulation-analogous to that of the Aharonov-Bohm effect resulting from the presence of a nonzero magnetic flux. In our experiments, the tunneling geometric phase is manifested through the interference of the corresponding supermodes. In this system, and for specific values of the twist rate, the tunneling between opposite cores ceases, thus signifying an Aharonov-Bohm suppression of tunneling. Our work provides the first observation of this intriguing effect in an optical setting.**

Quantum tunneling plays a central role in a number of intriguing phenomena in physics[1–5]. An archetypical manifestation of this quantum effect is the possibility of electron tunneling between multiple quantum wells when separated by potential barriers. Interestingly, on many occasions, even this simple class of systems can exhibit some unexpected quantum behaviors. For instance, complete quenching of coherent quantum tunneling can be achieved by appropriately imposing a time-harmonic drive to a double-well potential or other more complex lattice systems[6–8]. Another intriguing process is that arising from the topological restoration of energy degeneracies associated with bound states. These effects can reveal themselves in multiple quantum wells that are arranged on a closed loop, when threaded by a constant magnetic flux of certain quantized magnitudes[9]. This occurs due to the interference between degenerate electronic wavefunctions, when a nonzero Aharonov-Bohm (AB) phase is accumulated[10,11]. However, despite early theoretical efforts, an experimental observation of this latter effect is still lacking, as of yet, within the context of solid-state physics - primarily due to practical challenges associated with the need for ultrahigh magnetic fluxes.

In recent years, the realization of synthetic gauge fields in physical settings involving neutral entities has provided a flexible platform to observe dynamics similar to those arising from the interaction of electrons with magnetic fields[12–14]. In general, such artificial magnetism can be achieved through either dynamic temporal/spatial modulation of the couplings[15–17], or via the use of geometric phases[18,19]. So far, this artificial magnetism has led to the demonstration of photonic topological insulators in both passive [17,20–22] and active arrangements[23,24], Landau levels for photons[25,26], as well as quantum many-body effects[27,28]. An important advantage offered by such schemes is the possibility for experimental demonstrations of a certain class of quantum phenomena without invoking magnetic fields, something that would have been otherwise impossible within the context of condensed matter physics. In optics, perhaps the earliest demonstration of a geometric phase is that associated with polarized light[29]. This arises when the polarization state of light follows a closed contour on the Poincaré sphere. As a result, the corresponding electric field amplitude acquires a geometric phase, known as the Pancharatnam-Berry phase[29,30]. Similar effects can also occur for linearly polarized light[31] or speckle patterns in a multimode optical fiber[32,33], when the direction of propagation varies in space.

In this article, we experimentally demonstrate an optical geometric phase which is associated with photon tunneling in a twisted multicore fiber structure. We show that this form of geometric phase results from a constant rotation in the local frame of the fiber, and appears in a chiral manner in the corresponding coupling coefficients between adjacent cores. Thus far, twisted photonic crystal fibers have been employed to demonstrate coreless light guiding, optical activity, as well as orbital angular momentum conservation[34–36]. Alternatively, in our study, we employ the twist-induced geometric phase to realize a synthetic magnetic field for the photon coupling between nearest-neighbor light channels. As proposed in recent studies[37,38], this is analogous to the Aharonov-Bohm effect associated with electron dynamics in the presence of a magnetic flux. We further exploit this analogy to demonstrate the Aharonov-Bohm suppression of light tunneling in a four-core twisted optical fiber. While this latter effect was originally predicted in the context of quantum electronics [9], its experimental observation has so far remained elusive due to the requirement of strong magnetic fields. In this regard, it is only recently that an experimental observation of this type has been reported in a system involving ultracold Ca ions in a linear Paul trap[39]. In our experiments, we investigate the effect of different twist rates, emulating synthetic gauge fields of varying magnitudes. In this respect, the conditions for a complete tunneling inhibition are investigated, both theoretically and experimentally. Our experimental results are in good agreement with those expected from theory for different twist rates. Moreover, we study the behavior of this same arrangement under high power conditions - where nonlinear effects start to antagonize the coupling mechanisms in the multicore system. In this highly nonlinear regime, we find that the suppression of tunneling still persists – a direct byproduct of the topological nature of the AB geometric phase. Finally, even in the case where each core is multimoded, we demonstrate that the AB inhibition of tunneling occurs in a universal fashion. In other words, this same process can totally eliminate the coupling for all higher-order modes.

# Results

### A twisted multicore fiber platform for realizing synthetic magnetic fields for photons

In order to demonstrate AB suppression of light tunneling because of geometric phase effects, we fabricated a four-core optical fiber structure as shown in the inset of Fig. 1. Each of the four circular cores is coupled to its nearest neighbors, while a fluorine-doped refractive index depression in the middle of the structure eliminates any cross-channeling effects between opposite cores (see Methods). An artificial gauge field is then imposed on this system after twisting the multicore fiber. In this case, the evolution of the optical modal field amplitudes is described by a paraxial wave equation, expressed in the twisted local frame, in a way analogous to that associated with electron wavefunctions in the presence of a magnetic field[19,37,40] (Fig. 1). In quantum mechanics, in the presence of a uniform magnetic field, a gauge transformation through Peierls substitution is known to reduce the electron dynamics to those expected under conventional zero-field conditions. This result can be directly extended in more complicated settings such as for example atomic lattices. By treating this configuration within the tight-binding formalism, one can then show that the magnetic field now manifests itself in the form of complex coupling coefficients, having phase factors given by the Peierls integral[41]. Similarly, in a twisted multicore fiber configuration, the coupling coefficients appearing in the coupled mode equations are accordingly modified as $\kappa_{mn}e^{i\varphi_{mn}}$, where $\kappa_{mn}$ represents the coupling or tunneling strength between subsequent cores in the absence of twisting, while the pertinent phase factor is given by:

$$\varphi_{mn} = k_0 \int_{\vec{r}_m}^{\vec{r}_n} \frac{1}{2}\vec{r} \times \vec{B}_{eff} \, . \overrightarrow{dl} \, , \qquad (1)$$

where $\vec{B}_{eff} = -2n_0\epsilon\hat{z}$ is the effective magnetic field induced by twisting, and $\epsilon = 2\pi/\Lambda$ is the angular twist rate. $\Lambda$ denotes the spatial pitch associated with this helical structure, $k_0$ represents the free-space wavenumber, $n_0$ is the refractive index of the cladding region, and $\vec{r}_{m,n}$ stands for the positions of the neighboring cores $m, n$. Under these conditions, the evolution of the modal field amplitudes $E_n$ within the cores are given by $id|\psi\rangle/dz + H|\psi\rangle = 0$, where $|\psi\rangle = [E_1, E_2, E_3, E_4]^T$ represents a complex state vector whose evolution is subjected to a Hamiltonian $H$ that is given by:

$$H = \begin{pmatrix} \beta_1 & \kappa e^{-i\phi} & 0 & \kappa e^{i\phi} \\ \kappa e^{i\phi} & \beta_2 & \kappa e^{-i\phi} & 0 \\ 0 & \kappa e^{i\phi} & \beta_3 & \kappa e^{-i\phi} \\ \kappa e^{-i\phi} & 0 & \kappa e^{i\phi} & \beta_4 \end{pmatrix}. \qquad (2)$$

In Eq. (2), $\beta_n$ stands for the propagation constants of the individual cores, $\kappa$ is the magnitude of the nearest neighbor coupling coefficient, while $\phi = (k_0\, n_0\, \epsilon D^2)/2$ represents the Aharonov-Bohm tunneling phase introduced by the twist, when the core distance between successive sites is $D$, and $\epsilon = 2\pi/\Lambda$. In the four-core structure prepared, all cores are identical, and so are the respective propagation constants $\beta_n$. In this case, the $\beta$ terms can be readily eliminated from the evolution equations through a trivial gauge transformation. As we will see however, this is no longer valid under nonlinear conditions.

## Controlling and suppressing optical tunneling via twist-induced geometric phases

Equation (2) clearly suggests that photon tunneling in a twisted multicore fiber arrangement is accompanied by a geometric phase accumulation, akin to that expected in electron dynamics from a path-dependent AB phase. In order to experimentally observe the $\phi$ phase in the four-core fiber structure, we initially excite core #1 with coherent light from an external cavity laser, operating at $\lambda = 1550\,nm$ with an output power of $\sim 1\,mW$. The coupling length between successive cores is estimated to be $L_c \approx 9\,cm$. We then monitor the optical power coupled to the opposite core #3, as we vary the magnitude of the effective magnetic field (resulting from different twist rates). At the same time, the power levels in cores #2 and #4 are also recorded. It is important to note that the multicore fiber used here is designed in such a way that at $\lambda = 1550\,nm$ only the fundamental $LP_{01}$ mode is supported by each of the individual cores (Methods). Figure 2 shows experimental results where the output intensity from core #3 is plotted against the AB phase $\phi$. In this same figure, the expected theoretical behavior (see Methods) as obtained after directly solving the dynamical evolution equations is also presented. These observations indicate that the AB-like suppression of light tunneling from core #1 to #3 occurs when the gauge field corresponds to $\phi = \pi/4$. This is formally analogous to AB suppression of tunneling for electrons in the presence of a specific magnetic flux. These results provide the first observation of this intriguing effect in an optical setting.

## Impact of nonlinearity on AB tunneling inhibition

We further explore how the aforementioned AB-like suppression of tunneling is affected by the Kerr nonlinearity of these silica multicore fibers. In this respect, we launch $\sim 400\,ps$ high intensity pulses at $\lambda = 1064\,nm$ from a Q-switched microchip laser into core #1. At this wavelength, the fiber cores in our structure support $LP_{11}$ modes in addition to the fundamental $LP_{01}$. Because of mode confinement, the $LP_{11}$ modes are very strongly coupled at $1064\,nm$, while the $LP_{01}$ are virtually uncoupled. The initial ratio between the powers launched in $LP_{01}$ and $LP_{11}$at core #1 is adjusted by cleaving the input facet of the fiber at an angle. In this experiment, the sole purpose of exciting the fundamental $LP_{01}$ mode is to introduce a variable "energy" detuning $\Delta\beta_{NL} = k_0 n_2 |E_{max}|^2$ in the cores – thus allowing us to study how the nonlinearity $n_2$ impacts the inhibition of tunneling dynamics of the $LP_{11}$ modes. This is achieved through cross-phase modulation effects in each core. To observe these effects, we performed intensity measurements both at low $\sim 500\,W$ and high $\sim 6\,kW$ power levels (Methods). Figure 3 depicts the output intensity profiles at the end of the four-core fiber, for different twist-induced phases $\phi$. As indicated by these results, at $\phi = \pi/4$, the AB-like inhibition of tunneling between opposite cores still takes place regardless of the optical power exciting this system. In agreement with previous theoretical studies, these observations suggest that this process remains unaffected even under highly nonlinear conditions[38]. This robustness can be understood through a formal perturbation analysis (Supplementary), indicating that to first-order any $\Delta\beta$ variations (linear or nonlinear) within the four cores do not affect the twist-induced degeneracy between the two groups of supermodes at $\phi = \pi/4$. This faithful tunneling inhibition is attributed to the topological nature of this optical Aharonov-Bohm phase. Finally, as the optical power injected in the first core increases, a discrete soliton forms around this waveguide channel - further suppressing any tunneling of light to the nearby cores, as also evident in Fig. 3.

**Tunneling suppression in higher-order spatial modes**

Of interest would be to also investigate the universality of this class of effects even in multimode environments. To do so, we used a CW input excitation at $\lambda = 665\ nm$ where each core can now support four different sets of modes ($LP_{01}, LP_{11}, LP_{21}, LP_{02}$). The input power used was $\sim 2\ mW$ so as to ensure linear conditions. For this set of parameters, the highest-order propagating mode incited in this system was $LP_{02}$. Our simulations indicate that while the coupling coefficients associated with this mode are significant, all other modes exhibit negligible couplings, at least for length scales involved in our experiment. This is in agreement with experimental observations (Methods). To study the prospect for Aharonov-Bohm inhibition of tunneling in this multimode case, we again excite the first core while we record the output intensity patterns corresponding to different values of $\phi$. These results are summarized in Fig. 4, where it is clearly evident that the tunneling suppression always occurs, regardless of the number of the modes involved in each individual waveguide element.

## Discussion

According to the Aharonov-Bohm effect electron beams acquire a path-dependent phase in their corresponding wavefunctions in the presence of a nonzero magnetic flux[10,11]. This phase shift is not limited to conducting electrons, but also arises within the context of quantum tunneling[9]. Even though the AB effect has been observed for electrons in a superconducting or conductive platform[42–44], its observation in quantum well tunneling settings has so far remained out of reach, mainly due to the demand for high magnetic fluxes- which are experimentally inaccessible. Quite recently, a relevant observation has been reported[39], where a linear Paul trap was used to establish a bistable potential with two degenerate eigenstates for Ca ions. A magnetic field was then applied to the structure, and the tunneling rate between the corresponding degenerate states was found to be affected by the associated magnetic flux.

In optics, as originally shown by Pancharatnam, a cyclic change in the polarization of a light beam will in general lead to a phase shift accumulated by its corresponding electric field amplitude[29]. This can be viewed as the equivalent of the AB phase, where the magnetic flux is now replaced by the solid angle subtended by the corresponding cyclic curve on the Poincaré sphere for propagating photons[30]. On the other hand, the structure implemented here in our study is analogous to the AB effect for tunneling electrons, as predicted in recent theoretical studies[38]. Here the uniform twist along a circular multi-core fiber acts as a synthetic magnetic field for photon-tunneling between adjacent cores. Accordingly, the magnitude of this gauge field can be conveniently varied through mechanical twisting of the fiber, until a complete suppression of coupling between the two opposite cores can be achieved.

As indicated by our experimental results (Fig. 3), the AB tunneling phase in the twisted multicore structure remains invariant regardless of the presence of optical nonlinearity. Moreover, as confirmed by the observations presented in Fig. 4, this same effect happens in a universal manner even for higher-order modes in our optical fiber platform. We would like to emphasize here that local defects in the individual waveguide channels, something inevitable in any experimental realization, would not significantly affect our results[38]. In other words, this AB-induced tunneling inhibition effect happens to be robust against perturbations. This is attributed to the topological nature of the AB phase, as can be confirmed by perturbation analysis (see Supplementary). Our observations suggest that similar twisted fiber systems can be envisioned as viable platforms for studying effects akin to those arising from the interaction of electrons with magnetic fields-especially in connection with topological phenomena.

## Methods

### Four-core optical fiber platform

For our experimental demonstrations, we fabricated a silica fiber consisting of four coupled cores, each with a diameter of $\sim 7.5\ \mu m$ and a numerical aperture of $NA = 0.12$. The neighboring elements were separated from each other by a distance of $D = 23\ \mu m$. To observe the tunneling suppression between opposite cores, it is essential that any cross couplings are suppressed so that the light propagation dynamics in the system are governed by nearest-neighbor interactions. To achieve this, we judiciously incorporated a fluorine-doped low-index core in the center of the fiber, having a diameter of $\sim 5\ \mu m$. In the absence of any twist, when core #1 is initially excited, the light intensity in cores #1 and #3 varies along the propagation distance $z$ according to $I_1(z) = \cos^4 \kappa z$ and $I_3(z) = \sin^4 \kappa z$, as obtained after solving the dynamical modal evolution equations when $\phi = 0$. In other words, light tends to tunnel between these two waveguide channels through cores #2 and #4, in a way similar to tunneling of electrons in a multi-well potential arranged on a circular geometry.

### Twisting the structure and the dependency of the coupling on the synthetic gauge field

In order to introduce an artificial "magnetic field" in our arrangement, we twisted the 4-core fiber. The fiber was excited at $\lambda = 1550\ nm$, with an external cavity laser with a CW output power of $\sim 1 mW$. The output intensity profile from the four cores was then recorded on a CMOS-based IR beam profiling camera. Moreover, to discern different cores in the structure, we used a visible imaging camera in order to capture both the input and output facets of the fiber, once it was twisted. In general, for a given twisting rate (corresponding to an AB phase of $\phi$), the light intensity in cores #1 and #3 is explicitly given by the expressions:

$$I_{1,3}(L) = \frac{1}{4}[\cos(2\kappa L \cos\phi) \pm \cos(2\kappa L \sin\phi)]^2, \qquad (3)$$

where $L$ is the length of the four-core fiber. Figure 2 depicts these theoretical results along with experimentally observed values for different values of $\phi$ when $L = 24\ cm$. As evident from Eq. (3), for the specific case of $\phi = \pi/4$, the third core will always remain dark, irrespective of the length $L$, in agreement with our experimental results.

### High power characterization and multimode behavior

For the nonlinear experiments, we used a Q-switched microchip laser emitting high intensity pulses of duration $\sim 400\ ps$ at a rate of $500\ Hz$, at $\lambda = 1064\ nm$. At this wavelength, the highest-order mode supported by the cores happens to be the $LP_{11}$(see Supplementary). Our analysis shows that this higher order mode strongly couples nearest neighbor elements with a coupling length of $L_c \approx 2.5\ cm$, while the fundamental mode remains nearly uncoupled. As indicated in the Supplementary, at high peak powers ($\sim 6\ kW$), the Kerr nonlinearity associated with silica in the fiber structure results in a detuning in the propagation constant of the excited core. This strong nonlinearly induced detuning starts to compete with the coupling effects, eventually forming discrete solitons in the excited core. This antagonizing effect of the nonlinearity with respect to light coupling is also evident in our experiments (Fig. 3). Finally, for $\lambda = 665\ nm$, each waveguide is multimoded and the mode with the highest coupling happens to be the $LP_{02}$, exhibiting $L_c \approx 10\,cm$ (see Supplementary). This is also evident in the experimental results where the intensity profile of the light coupled is clearly that corresponding to a radially symmetric $LP_{02}$ mode.

## Acknowledgements

This effort was partially supported by the Office of Naval Research (ONR) (MURI N00014-17-1-2588, N0001416-1- 2640, N00014-18-1-2347), National Science Foundation (NSF) (ECCS-1711230, ECCS 1454531, DMR- 1420620, ECCS 1757025), HEL-JTO (W911NF-12-1-0450), Army Research Office (ARO) (W911NF-12-1-0450), Air Force Office of Scientific Research (AFOSR) (FA955015-10041), Army Research Office (ARO) (W911NF-17-1-0481) and Qatar National Research Fund (QNRF) (NPRP 9-020-1-006).

## Author Contributions

M.P., H.L.A., M. K., R.A.C. and D.N.C. conceived the experiments presented here. J.E.A.L., R.A.C. fabricated the fiber structure. M.P. and H.L.A. built the optical setup and performed the measurements. R.A.C., M.K., and D.N.C. supervised the project. All authors contributed equally to the writing of this manuscript.

**Figures and figure captions**

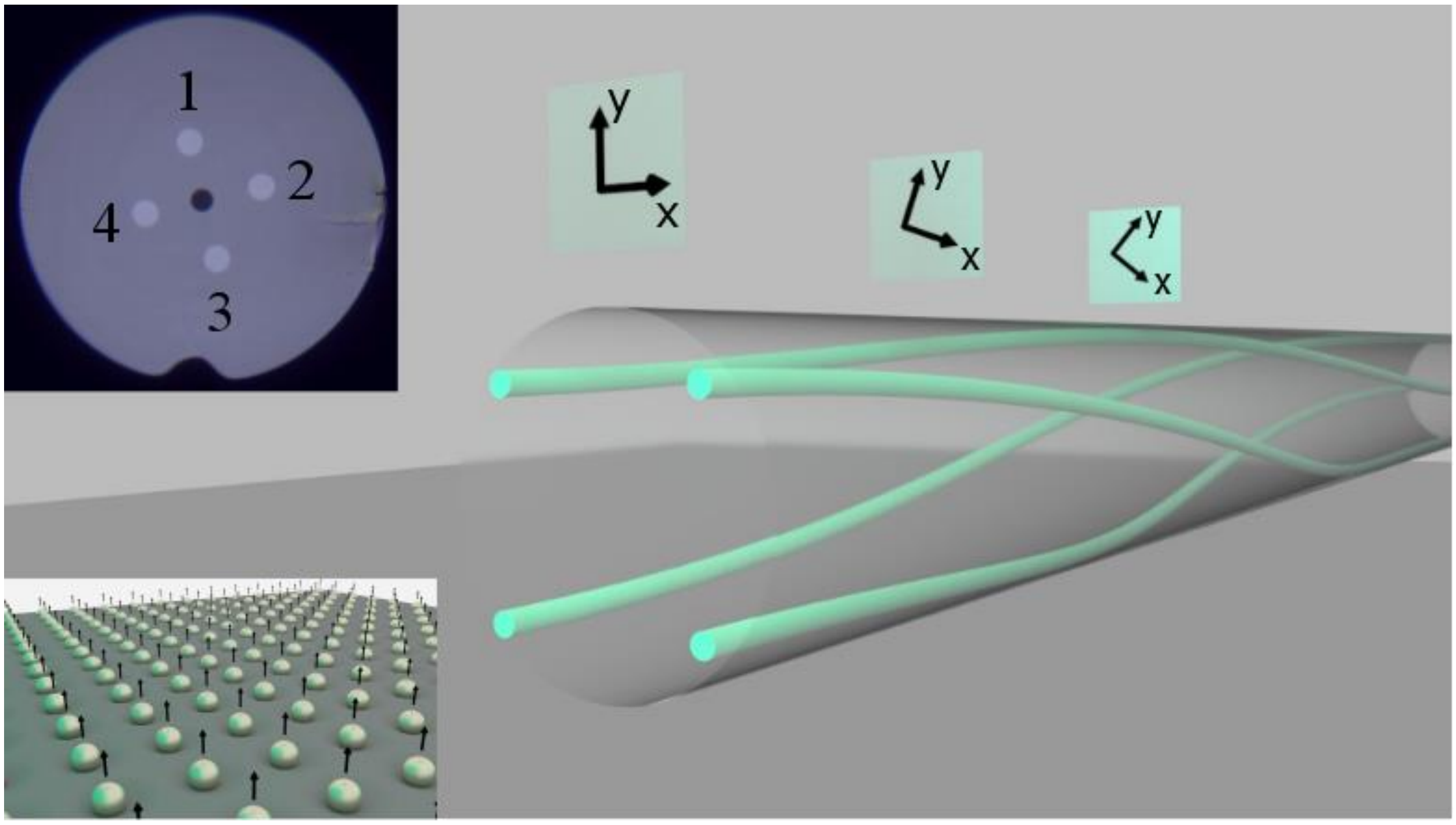

**Figure 1 | Twisted fiber structures as a platform for realizing synthetic magnetic fields for photons.** A twisted four-core optical fiber in which the photon tunneling evolution dynamics are analogous to those expected from electrons in the presence of a magnetic field. The constantly rotating local transverse coordinates are depicted at three different planes. The top inset shows a microscope image of the input facet of the four-core fiber used in our experiments. The low-index fluorine-doped core is clearly visible at the center of the fiber. The bottom inset depicts a schematic of a two-dimensional atomic lattice in the presence of a static perpendicular magnetic field (arrows), where a tight-binding formalism can be utilized to describe the electronic band structure after a Peierls substitution.

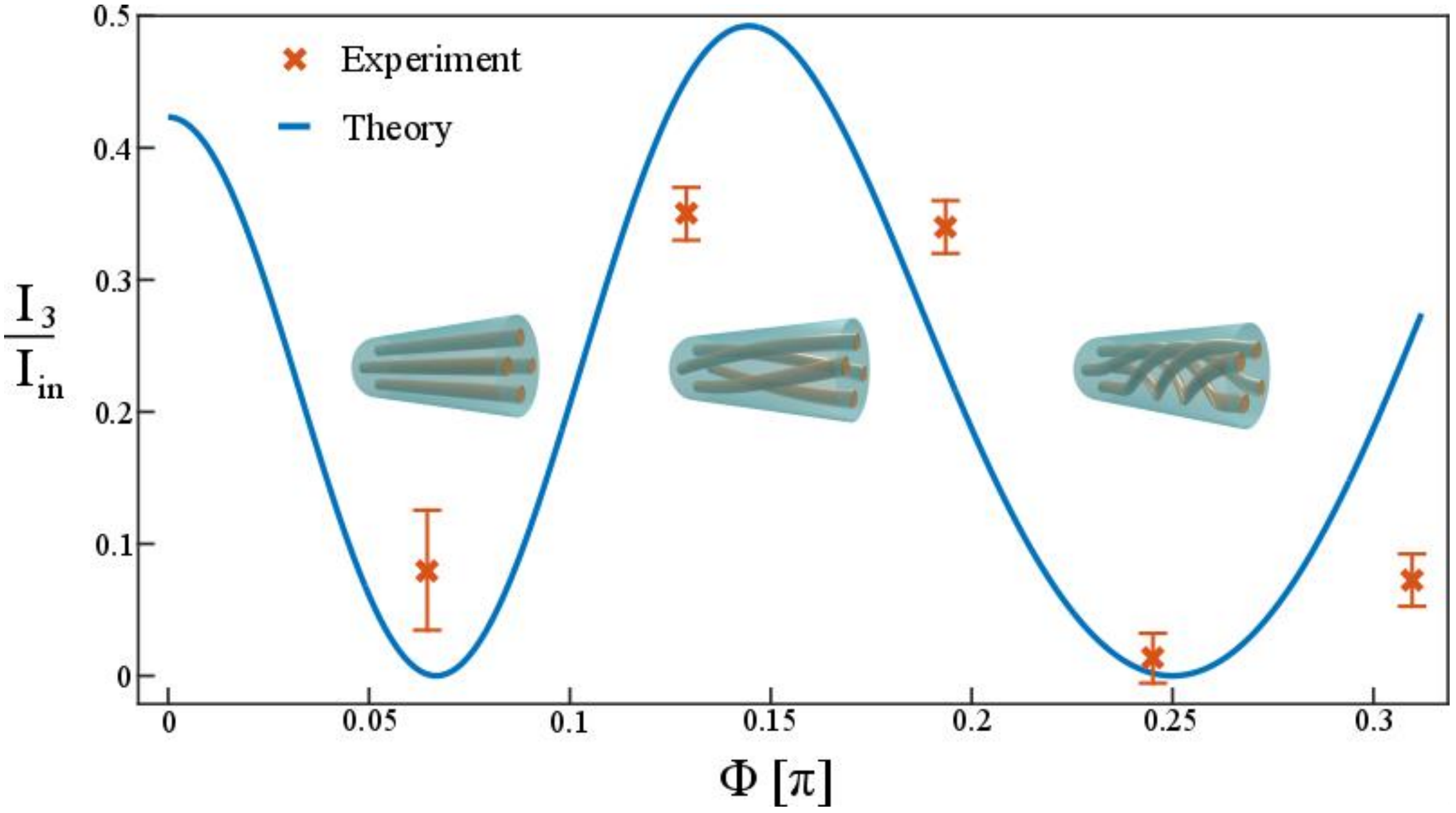

**Figure 2 | Dependence of optical tunneling dynamics on the AB phase.** Normalized light intensity at the output of core #3 for different values of the AB phase $\phi$ (as induced by different twist rates). In all cases, core #1 is excited at the input with CW laser light at $\lambda = 1550\,nm$. Theoretical results corresponding to the same set of parameters are also provided for comparison. At $\phi = \pi/4$, the third core always remains dark, clearly indicating AB tunneling suppression.

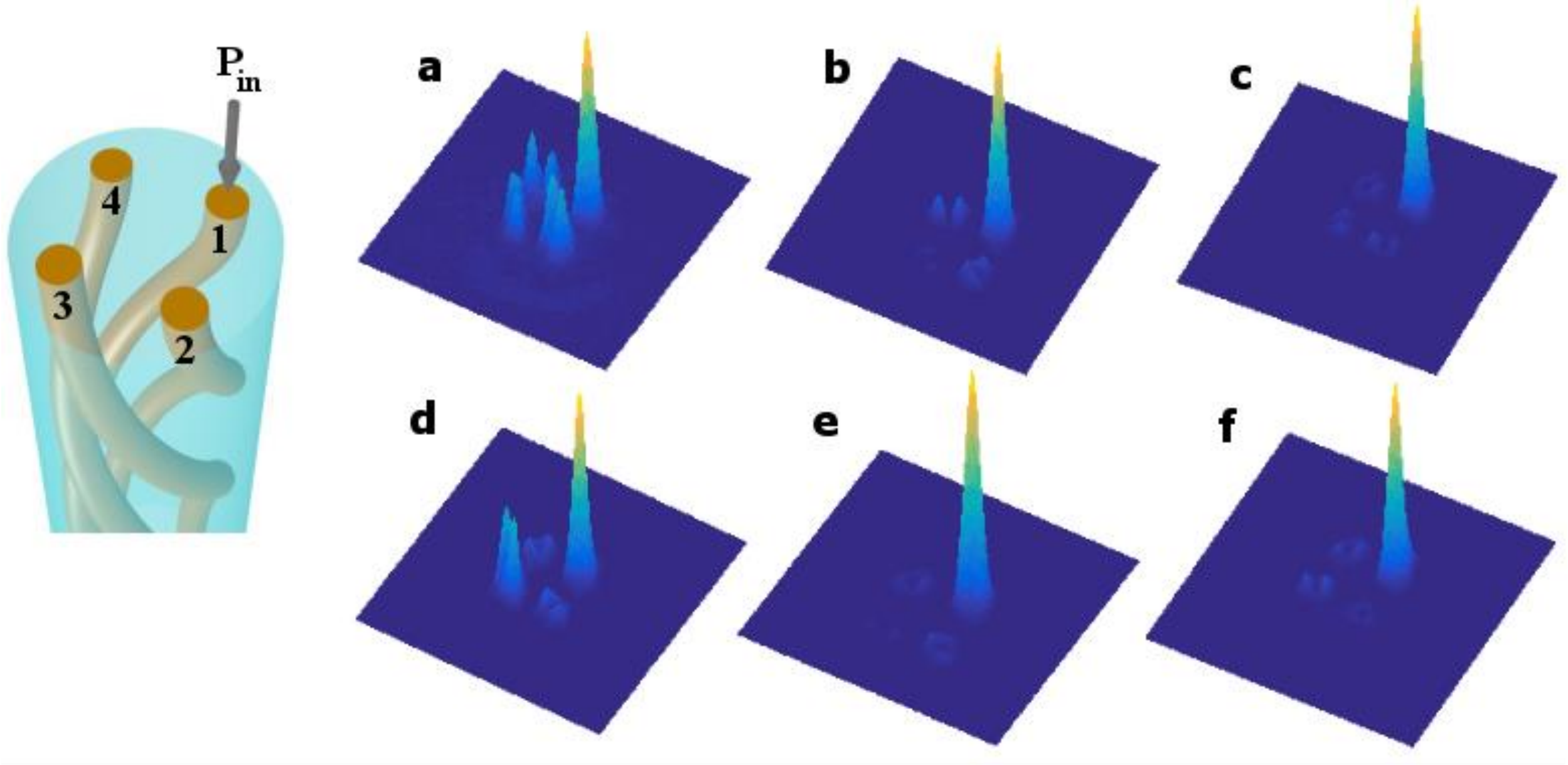

**Figure 3 | AB inhibition of tunneling in the presence of optical nonlinearities.** Output light intensity profiles from a twisted, 24 cm long, four-core fiber when only core #1 is quasi-linearly excited for **a**, $\phi = 0$ (no twist), **b**, $\phi = \pi/4$, and **c**, $\phi \approx 0.27\pi$. In (**a**-**c**), the pulses used had a peak power $\sim 500W$ at $\lambda = 1064\, nm$. Plots (**d**-**f**) show similar results when the input peak power is $\sim 6kW$ and hence nonlinear Kerr effects are at play. It is evident that the coupling between cores #1 and #3 is completely suppressed in both cases (**b**,**e**), regardless of the power levels used, indicating an immunity of the AB tunneling suppression against nonlinear index changes. For higher input powers (**d**-**f**), the self-focusing nonlinearity further suppresses light coupling, even among adjacent cores. The inset on the left depicts the relative arrangement of the four cores in this particular experiment.

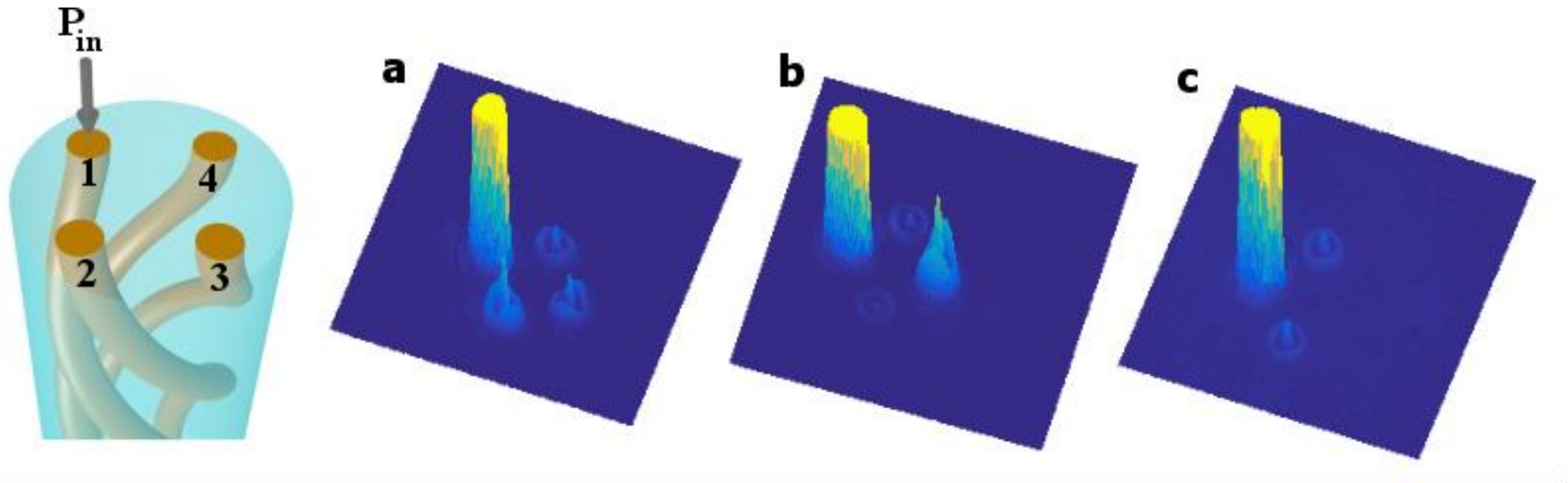

**Figure 4 | AB tunneling suppression for higher-order modes.** Light intensity distributions at the output of a twisted four-core fiber when the higher-order $LP_{02}$ mode is excited with CW light at $\lambda = 665\,nm$. These results are presented for **a**, $\phi = 0$ (no twist), **b**, $\phi \approx 0.11\pi$, and **c**, $\phi = \pi/4$. Even though in the excited core #1, most of the optical power resides in the fundamental $LP_{01}$ mode, only the $LP_{02}$ mode appears in the remaining cores due to its higher coupling coefficient. Plot **c** clearly reveals that AB suppression of light tunneling occurs in a universal fashion, even for higher-order modes. The inset on the left depicts the arrangement of the cores corresponding to these observations.

# Supplementary Materials for

## Observation of twist-induced geometric phases and inhibition of optical tunneling via Aharonov-Bohm effects

Midya Parto, Helena Lopez-Aviles, Jose E. Antonio-Lopez, Mercedeh Khajavikhan, Rodrigo Amezcua-Correa, and Demetrios N. Christodoulides

CREOL, The College of Optics & Photonics, University of Central Florida, Orlando, FL 32816, USA.

**This file includes:**

# 1. Supermodes of twisted multicore optical fibers

In this section, we provide analytical derivations for the supermodes of a twisted multicore optical fiber, focusing on the effect of the geometric phase on the eigenstates and their corresponding eigenvalues.

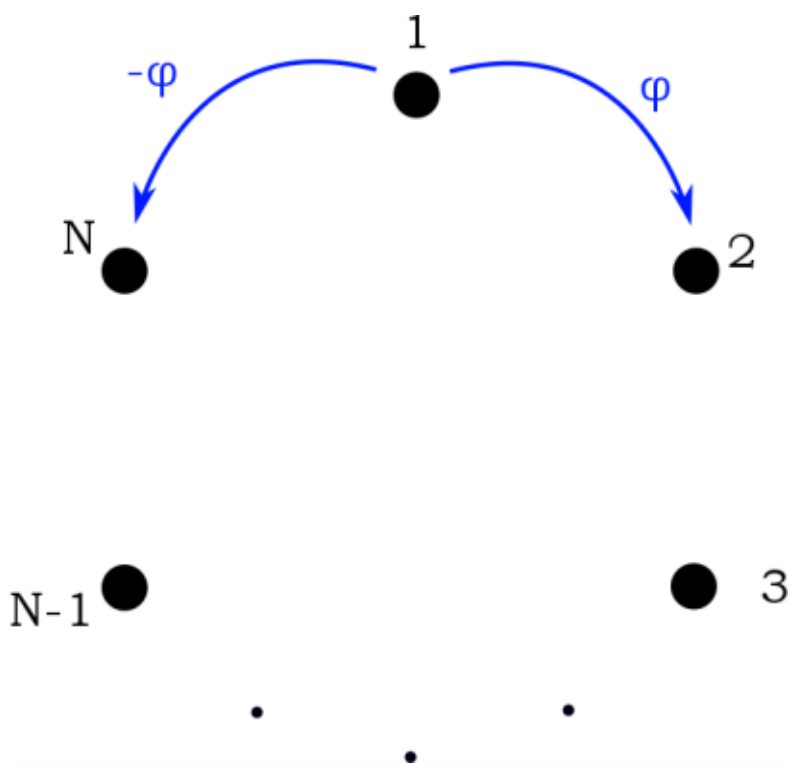

**Fig. S1. Twisted N-core fiber.** The tunneling Aharonov-Bohm phase is depicted in the figure by $\varphi$.

As discussed in the main text, light dynamics in a twisted multicore optical fiber can be described by the following eigenvalue equation:

$$id|\psi\rangle/dz + H|\psi\rangle = 0,$$

$$H = \begin{bmatrix} \beta_1 & \kappa e^{i\phi} & \cdots & 0 & \kappa e^{-i\phi} \\ \kappa e^{-i\phi} & \beta_2 & & 0 & 0 \\ \vdots & & \ddots & \vdots & \\ 0 & 0 & \cdots & \beta_{N-1} & \kappa e^{i\phi} \\ \kappa e^{i\phi} & 0 & & \kappa e^{-i\phi} & \beta_N \end{bmatrix}, \quad (1)$$

where $|\psi\rangle = [E_1, E_2, \ldots, E_N]^T$ represents the electric field amplitudes of an eigenstate of the system. Here we assume that all the cores are similar, hence $\beta_i = 0$. In this case, the general form of the eigenstates is given by:

$$E_n = Ae^{i\mu z}e^{iQn}, \quad (2)$$

where $\mu$ represents the eigenvalues, while $Q$ denotes the corresponding Bloch momenta of the eigenstates. Substituting this into the eigenvalue equation, one obtains the following form for the eigenvalues:

$$\mu = 2\kappa\cos(Q - \phi). \quad (3)$$

Applying periodic boundary conditions results in quantized Bloch momenta $Q = 2\pi m/N, m = 0,1, \ldots, N-1$. It is evident that the tunneling Aharonov-Bohm phase changes the eigenvalues in this system through shifting the Bloch momentum $Q$, while it does not affect the form of the eigenstates. This is similar to the way a static magnetic field would modify the electronic states in a periodic 2D arrangement.

By setting $N = 4$, the eigenstates associated with the twisted four-core fiber considered in our experiments can be obtained as:

$$|\psi_0\rangle = \frac{1}{2}\begin{pmatrix}1\\1\\1\\1\end{pmatrix}, |\psi_1\rangle = \frac{1}{2}\begin{pmatrix}1\\i\\-1\\-i\end{pmatrix}, |\psi_2\rangle = \frac{1}{2}\begin{pmatrix}1\\-1\\1\\-1\end{pmatrix}, |\psi_3\rangle = \frac{1}{2}\begin{pmatrix}1\\-i\\-1\\i\end{pmatrix}, \quad (4)$$

while their respective eigenvalues are:

$$\mu_0 = 2\kappa\cos\phi\,, \mu_1 = 2\kappa\sin\phi\,, \mu_2 = -2\kappa\cos\phi\,, \mu_3 = -2\kappa\sin\phi. \quad (5)$$

For a twist rate corresponding to $\phi = \pi/4$, these will form two pairs of degenerate eigenmodes $\mu_0 = \mu_1 = \sqrt{2}\kappa$ and $\mu_2 = \mu_3 = -\sqrt{2}\kappa$. In this case, if core #1 is excited at the input, the initial state of the system at $z = 0$ can be expanded in terms of the given eigenstates as:

$$|\psi_{in}\rangle = \begin{pmatrix}1\\0\\0\\0\end{pmatrix} = \frac{1}{2}\sum_{i=0}^{3}|\psi_i\rangle. \quad (6)$$

Therefore, the propagated state at an arbitrary distance $z$ will then be:

$$|\psi(z)\rangle = \frac{1}{2}\sum_{i=0}^{3} e^{i\mu_i z}|\psi_i\rangle = \frac{1}{2}\begin{pmatrix}\cos(2\kappa z\cos\phi) + \cos(2\kappa z\sin\phi)\\ i\sin(2\kappa z\cos\phi) - \sin(2\kappa z\sin\phi)\\ \cos(2\kappa z\cos\phi) - \cos(2\kappa z\sin\phi)\\ i\sin(2\kappa z\cos\phi) + \sin(2\kappa z\sin\phi)\end{pmatrix}, \quad (7)$$

Substituting $\phi = \pi/4$, we find:

$$|\psi(z)\rangle = \frac{1}{2}\begin{pmatrix}2\cos(\sqrt{2}\kappa z)\\ (i-1)\sin(\sqrt{2}\kappa z)\\ 0\\ (i+1)\sin(\sqrt{2}\kappa z)\end{pmatrix}, \quad (8)$$

indicating that core #3 will always remain dark irrespective of the length of the fiber.

## 2. Perturbation analysis of the tunneling inhibition

In this section, we consider how small perturbations in terms of detuning of individual cores or variations in coupling coefficients between nearby cores would affect the AB tunneling inhibition effect in our four-core twisted fiber.

We first consider a diagonal perturbation in the Hamiltonian describing the structure:

$$H' = H + \Delta H_1,$$

$$H = \begin{pmatrix} 0 & \kappa e^{-i\phi} & 0 & \kappa e^{i\phi} \\ \kappa e^{i\phi} & 0 & \kappa e^{-i\phi} & 0 \\ 0 & \kappa e^{i\phi} & 0 & \kappa e^{-i\phi} \\ \kappa e^{-i\phi} & 0 & \kappa e^{i\phi} & 0 \end{pmatrix},\ \Delta H_1 = \begin{pmatrix} \epsilon_1 & 0 & 0 & 0 \\ 0 & \epsilon_2 & 0 & 0 \\ 0 & 0 & \epsilon_3 & 0 \\ 0 & 0 & 0 & \epsilon_4 \end{pmatrix}. \quad (9)$$

In this case, the first-order approximation of the perturbed eigenvalues will be:

$\mu'_i \approx \ \mu_i + \langle \psi_i | \Delta H_1 | \psi_i \rangle = \mu_i + \Delta\mu_i,$

$$\Delta\mu_i = \frac{1}{2}\begin{pmatrix} 1 & e^{iQ} & e^{i2Q} & e^{i3Q} \end{pmatrix} \begin{pmatrix} \epsilon_1 & 0 & 0 & 0 \\ 0 & \epsilon_2 & 0 & 0 \\ 0 & 0 & \epsilon_3 & 0 \\ 0 & 0 & 0 & \epsilon_4 \end{pmatrix} \frac{1}{2} \begin{pmatrix} 1 \\ e^{-iQ} \\ e^{-i2Q} \\ e^{-i3Q} \end{pmatrix} = \frac{1}{4}\sum_{i=0}^{3} \epsilon_i, \quad (10)$$

This shows that a diagonal perturbation would lead into the same first-order correction for all the four eigenvalues. Therefore, to first-order, we still have two pair of degenerate eigenstates and hence the tunneling inhibition is preserved.

Now let us consider a perturbation in the coupling coefficients:

$$H' = H + \Delta H_2,\ \Delta H_2 = \begin{pmatrix} 0 & \delta\kappa_{12} & 0 & \delta\kappa_{14} \\ \delta\kappa_{12}^* & 0 & \delta\kappa_{23} & 0 \\ 0 & \delta\kappa_{23}^* & 0 & \delta\kappa_{34} \\ \delta\kappa_{14}^* & 0 & \delta\kappa_{34}^* & 0 \end{pmatrix}. \quad (11)$$

Similar to the previous case, the first-order perturbations can be obtained as:

$$\Delta\mu_0 = \frac{1}{2}\begin{pmatrix} 1 & 1 & 1 & 1 \end{pmatrix} \Delta H_2 \frac{1}{2} \begin{pmatrix} 1 \\ 1 \\ 1 \\ 1 \end{pmatrix} = \frac{1}{2} Re\{\delta\kappa_{12} + \delta\kappa_{23} + \delta\kappa_{34} + \delta\kappa_{14}\},$$

$$\Delta\mu_1 = \frac{1}{2}\begin{pmatrix} 1 & -i & -1 & i \end{pmatrix} \Delta H_2 \frac{1}{2} \begin{pmatrix} 1 \\ i \\ -1 \\ -i \end{pmatrix} = -\frac{1}{2} Im\{\delta\kappa_{12} + \delta\kappa_{23} + \delta\kappa_{34} - \delta\kappa_{14}\},$$

$$\Delta\mu_2 = \frac{1}{2}\begin{pmatrix} 1 & -1 & 1 & -1 \end{pmatrix} \Delta H_2 \frac{1}{2} \begin{pmatrix} 1 \\ -1 \\ 1 \\ -1 \end{pmatrix} = -\frac{1}{2} Re\{\delta\kappa_{12} + \delta\kappa_{23} + \delta\kappa_{34} + \delta\kappa_{14}\},$$

$$\Delta\mu_3 = \frac{1}{2}\begin{pmatrix} 1 & i & -1 & -i \end{pmatrix} \Delta H_2 \frac{1}{2} \begin{pmatrix} 1 \\ -i \\ -1 \\ i \end{pmatrix} = \frac{1}{2} Im\{\delta\kappa_{12} + \delta\kappa_{23} + \delta\kappa_{34} - \delta\kappa_{14}\}. \quad (12)$$

We are particularly interested in perturbations in the magnitudes of the coupling coefficients, therefore we have:

$$\delta\kappa_{12} = |\delta\kappa_{12}|e^{-i\phi}, \delta\kappa_{23} = |\delta\kappa_{23}|e^{-i\phi}, \delta\kappa_{34} = |\delta\kappa_{34}|e^{-i\phi}, \delta\kappa_{14} = |\delta\kappa_{14}|e^{i\phi}, \quad (13)$$

Hence:

$$\Delta\mu_0 = \frac{1}{2}(|\delta\kappa_{12}| + |\delta\kappa_{23}| + |\delta\kappa_{34}| + |\delta\kappa_{14}|)\cos\phi,$$

$$\Delta\mu_1 = \frac{1}{2}(|\delta\kappa_{12}| + |\delta\kappa_{23}| + |\delta\kappa_{34}| + |\delta\kappa_{14}|)\sin\phi,$$

$$\Delta\mu_2 = -\frac{1}{2}(|\delta\kappa_{12}| + |\delta\kappa_{23}| + |\delta\kappa_{34}| + |\delta\kappa_{14}|)\cos\phi,$$

$$\Delta\mu_3 = -\frac{1}{2}(|\delta\kappa_{12}| + |\delta\kappa_{23}| + |\delta\kappa_{34}| + |\delta\kappa_{14}|)\sin\phi. \qquad (14)$$

As indicated by these results, we see that in general, arbitrary perturbations in the coupling strengths result in different corrections $\Delta\mu_i$. However, for the special case where $\phi = \pi/4$ which corresponds to the inhibition of the tunneling, one obtains $\Delta\mu_0 = \Delta\mu_1$ and $\Delta\mu_2 = \Delta\mu_3$, showing that the underlying degeneracy among the eigenstates in our system persists in the presence of this type of disorder. This in turn means that the AB tunneling inhibition itself is robust with respect to both diagonal and off-diagonal perturbations.

# 3. Coupling of the fundamental mode and higher-order modes

Here we discuss finite-element simulations we performed to determine coupling strengths of the fundamental mode as well as higher-order modes in different wavelengths of our experiments.

In the first set of our experiments, we used light at a wavelength of $\lambda = 1550\,nm$. In this case, the associated $V$ number of the individual cores is:

$$V = k_0 r_{core} NA = 1.8, \qquad (15)$$

and therefore each core is single-moded in this case ($LP_{01}$). As we decrease the wavelength to $\lambda = 1064\,nm$ and $\lambda = 665\,nm$, this $V$ number increases, leading to a multimode behavior within each of the light channels. Supplementary Table 1 summarizes these results, where the corresponding mode profiles together with their nearest-neighbor coupling coefficients are also reported.

| $\lambda = 1550\,nm$ | | |
|---|---|---|
| Mode | Coupling coefficient $\kappa\ [m^{-1}]$ | Mode profile |
| $LP_{01a}$ | 16 | 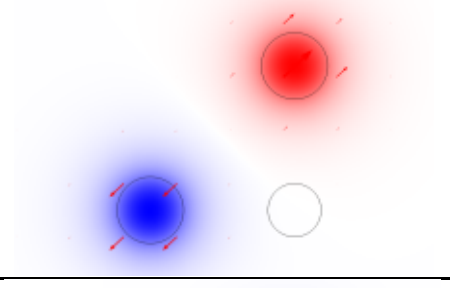 |
| $LP_{01b}$ | 16 | 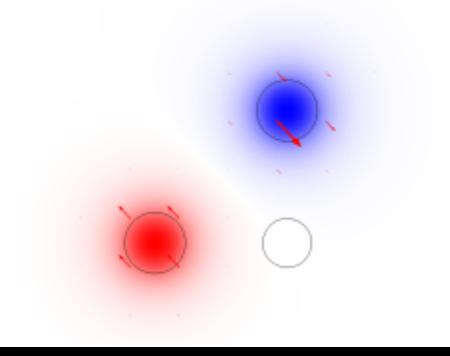 |
| $\lambda = 1064\,nm$ | | |
| $LP_{01}$ | 0.24 | 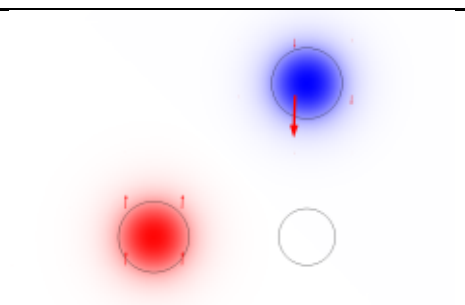 |
| $LP_{11a}$ | 63 | 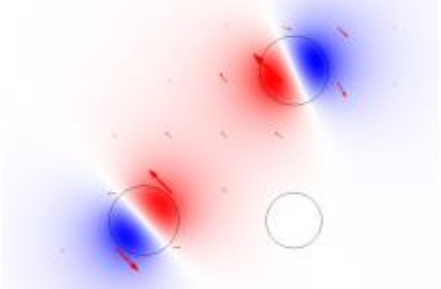 |
| $LP_{11b}$ | 61.5 | 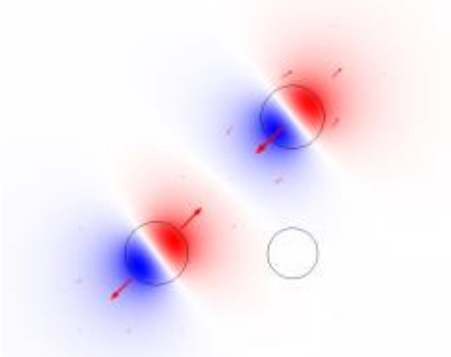 |
| $LP_{11c}$ | 10 | 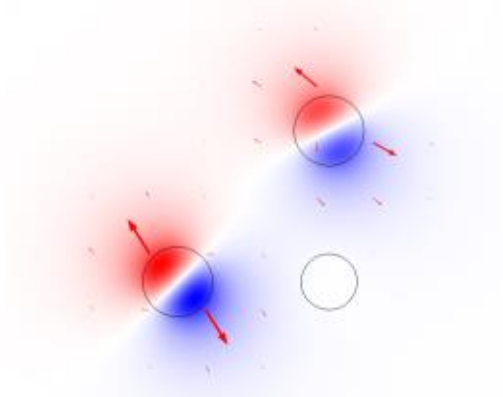 |
| $LP_{11d}$ | 11.5 | 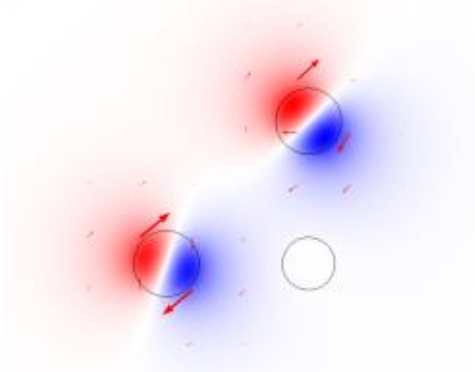 |
| | $\lambda = 665\ nm$ | |
| $LP_{01}$ | 0.01 | 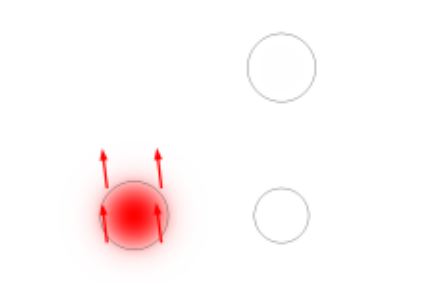 |
| $LP_{11}$ | 0.08 | 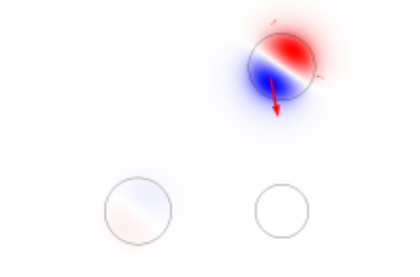 |
| $LP_{21}$ | 0.38 | 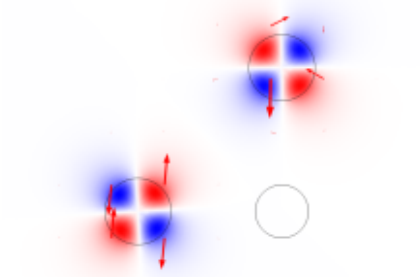 |
| $LP_{02a}$ | 16 | 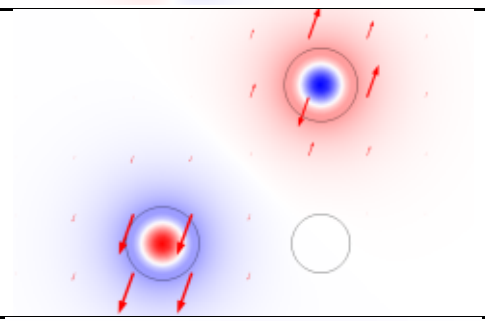 |
| $LP_{02b}$ | 16 | 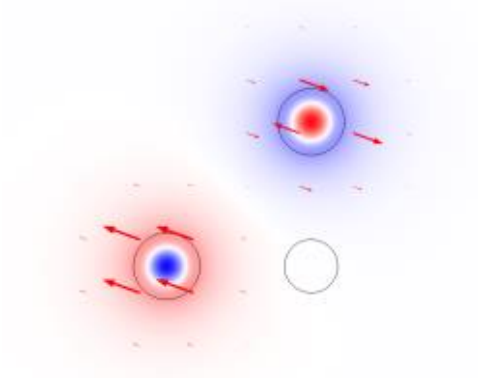 |

**Table S1. Multimode behavior of the designed four-core fiber.** Optical modes supported by each of the cores in different wavelength regimes and their corresponding coupling strengths. Here the transverse component of the electric field is depicted, with red and blue corresponding to positive and negative values, respectively. The arrows indicate the actual direction of the transverse electric fields of the associated supermodes.

As mentioned in the manuscript, to observe the tunneling suppression between opposite cores, it is essential that cross-coupling between them is highly suppressed so that the light propagation dynamics in the system are governed by nearest-neighbor interactions. In order to achieve this, we incorporated a fluorine-doped low-index core in the center of the fiber, having a diameter of ~5 μm. Our simulations show that in the absence of such measures the cross-coupling is around ~ 6% of the nearest-neighbor coupling coefficients, while by introducing the refractive index suppression it is reduced to ~ 2%.

# 4. Kerr induced detuning in high powers

In this section we consider the effect of high power excitation which leads into self-focusing detuning in our silica fiber platform.

As mentioned in the main text, our high power experiments were performed at $\lambda = 1064\ nm$. We used optical pulses of duration $\sim 400\ ps$ from a Q-switched microchip laser with peak powers of $\sim 500\ W$ and $\sim 6\ kW$. Therefore, Kerr induced detuning in the excited core #1 in each case will be:

$$\Delta\beta_{low} = k_0 n_2^I I_{low} \approx 4\ m^{-1} \tag{16}$$

$$\Delta\beta_{high} = k_0 n_2^I I_{high} \approx 48\ m^{-1}. \tag{17}$$

According to these, in the high optical power regime, the nonlinear induced detuning in the propagation constant of the excited core is comparable in magnitude with the coupling coefficient of the higher-order $LP_{11}$ mode. This results in a more confinement of the light in the excited core #1, further decreasing the coupling to the other cores.

# 5. Coupling suppression of higher-order modes

Here we consider coupled mode analysis of the twisted fiber for higher-order modes, e.g. the $LP_{11}$ mode arising in our nonlinear experiments. As indicated by our simulations in Table S1, the coupling between cores for higher-order modes can in general differ due to specific orientation of the corresponding mode. As shown in Fig. S2, this can be understood by noting that the coupling between nearby cores supporting e.g. $LP_{11}$ mode is directional, leading to different couplings $\kappa_1$ and $\kappa_2$ in this case ($\kappa_1 > \kappa_2$).

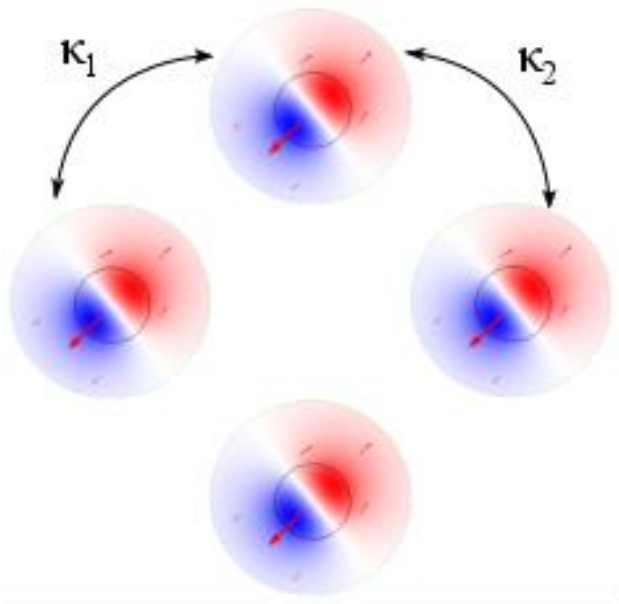

**Fig. S2. Inhomogeneous couplings among cores.** Different coupling coefficients among various nearby cores in a four-core system for an $LP_{11}$ mode.

Consequently, the coupled mode equations can be formulated as discussed in the main text, except that now the equivalent Hamiltonian H is given as

$$H_1 = \begin{pmatrix} 0 & \kappa_2 e^{-i\phi} & 0 & \kappa_1 e^{i\phi} \\ \kappa_2 e^{i\phi} & 0 & \kappa_1 e^{-i\phi} & 0 \\ 0 & \kappa_1 e^{i\phi} & 0 & \kappa_2 e^{-i\phi} \\ \kappa_1 e^{-i\phi} & 0 & \kappa_2 e^{i\phi} & 0 \end{pmatrix}. \quad (18)$$

For a twist rate corresponding to $\phi = \pi/4$, it can be shown that $H_1$ exhibits two pairs of degenerate eigenmodes $\mu_0 = \mu_1 = \mu_- = -\sqrt{\kappa_1^2 + \kappa_2^2}$ and $\mu_2 = \mu_3 = \mu_+ = \sqrt{\kappa_1^2 + \kappa_2^2}$, together with their associated eigenvectors:

$$|\psi_0\rangle = \frac{1}{2}\begin{pmatrix} 1 \\ -e^{-i(\frac{\pi}{4}+\theta)} \\ -1 \\ e^{-i(\frac{\pi}{4}+\theta)} \end{pmatrix}, |\psi_1\rangle = \frac{1}{2}\begin{pmatrix} 1 \\ -e^{-i(\frac{\pi}{4}-\theta)} \\ 1 \\ -e^{-i(\frac{\pi}{4}-\theta)} \end{pmatrix}, |\psi_2\rangle = \frac{1}{2}\begin{pmatrix} 1 \\ e^{-i(\frac{\pi}{4}+\theta)} \\ -1 \\ -e^{-i(\frac{\pi}{4}+\theta)} \end{pmatrix}, |\psi_3\rangle = \frac{1}{2}\begin{pmatrix} 1 \\ -e^{-i(\frac{\pi}{4}-\theta)} \\ 1 \\ e^{-i(\frac{\pi}{4}-\theta)} \end{pmatrix}, \quad (19)$$

where $\tan\theta = \kappa_1/\kappa_2$. Under these conditions, if core #1 is initially excited at the input, the initial state of the system at $z = 0$ can be expanded in terms of the given eigenstates as:

$$|\psi_{in}\rangle = \begin{pmatrix} 1 \\ 0 \\ 0 \\ 0 \end{pmatrix} = \frac{1}{2}\sum_{i=0}^{3} |\psi_i\rangle. \quad (20)$$

Therefore, the propagated state at an arbitrary distance $z$ will then be

$$|\psi(z)\rangle = \frac{1}{2}\sum_{i=0}^{3} e^{i\mu_i z}|\psi_i\rangle = \frac{1}{2}\begin{pmatrix} e^{i\mu_- z} + e^{i\mu_+ z} \\ -e^{-\frac{i\pi}{4}}\cos\theta\left(e^{i\mu_- z} - e^{i\mu_+ z}\right) \\ 0 \\ -ie^{-\frac{i\pi}{4}}\sin\theta\left(e^{i\mu_- z} - e^{i\mu_+ z}\right) \end{pmatrix}, \quad (21)$$

indicating that core #3 will always remain dark irrespective of the length of the fiber.